\documentclass[letterpaper,aps,prl,twocolumn,superscriptaddress,groupedaddress]{revtex4}  
\usepackage{graphicx}  
\usepackage{dcolumn}   
\usepackage{bm}        
\usepackage{amssymb,amsmath}   
\usepackage{array}

\makeatletter
\newcommand*{\balancecolsandclearpage}{%
  \close@column@grid
  \clearpage
  \twocolumngrid
}
\makeatother

\usepackage{color}

\begin{document}




\title{Evolution in range expansions with competition at rough boundaries}
\author{Sherry Chu}
\affiliation{Department of Physics, Massachusetts Institute of Technology, Cambridge, MA 02139, USA}
\author{Mehran Kardar}
\affiliation{Department of Physics, Massachusetts Institute of Technology, Cambridge, MA 02139, USA}
\author{David R.~Nelson}
\affiliation{Department of Physics, Department of Molecular and Cellular Biology and School of Engineering and Applied Sciences, Harvard University, Cambridge, MA 02138, USA}
\author{Daniel A.~Beller}
\affiliation{Department of Physics, University of California, Merced, CA 95343, USA}
\email{dbeller@ucmerced.edu}       
\date{\today}

\begin{abstract}
When a biological population expands into new territory, genetic drift develops an enormous influence on evolution at the propagating front. In such range expansion processes, fluctuations in allele frequencies occur through stochastic \textit{spatial} wandering of both genetic lineages and the boundaries between genetically segregated sectors. Laboratory experiments on microbial range expansions have shown that this stochastic wandering, transverse to the front, is superdiffusive due to the front's growing roughness, implying much faster loss of genetic diversity than predicted by simple flat front diffusive models. We study the evolutionary consequences of this superdiffusive wandering using two complementary numerical models of range expansions: the stepping stone model, and a new interpretation of the model of directed paths in random media, in the context of a roughening population front. Through these approaches we compute statistics for the times since common ancestry for pairs of individuals with a given spatial separation at the front, 
and we explore how environmental heterogeneities can locally suppress these superdiffusive fluctuations. 
\end{abstract}

\maketitle

\section{\label{sec:intro}Introduction}

In evolutionary biology, changes in an allele's frequency in a population are driven not only by Darwinian selection but also by random fluctuations, the phenomenon of genetic drift. Selectively neutral or even deleterious alleles can rise to prominence purely by chance. In many scenarios an individual competes directly only with a small subset of the population, e.g. due to spatial proximity, and this small effective population size increases the influence of genetic drift~\cite{korolev2010genetic}. 

Range expansions provide an important example: When a population expands spatially into new territory, as during species invasion or following environmental changes, the new territory is dominated by the descendants of a few ancestors at the expansion front. Genetic drift is amplified by the small effective population size at the  front~\cite{korolev2010genetic} -- the  founder effect -- and by the related phenomenon of gene ``surfing'', in which alleles that happen to be present at the front spread to high frequency in the newly occupied space, despite being selectively neutral or even deleterious ~\cite{hallatschek2007genetic,excoffier2008surfing}. 

Genetic drift in range expansions strongly ties fluctuations in allele frequencies to spatial fluctuations. In laboratory experiments, Hallatschek et al.~\cite{hallatschek2007genetic} have shown that microbial range expansions develop, after a short demixing time, genetic sectors containing almost exclusively the descendants of a single individual. Thereafter, genetic drift occurs through \textit{spatial} fluctuations of the sector boundaries, with a sector lost from the front each time two sector boundaries intersect. Similarly, the geneological ancestry tree traced backward in time from the front becomes a tree of space curves that fluctuate transversely to the front propagation direction and coalesce upon intersection~\cite{ELE:ELE12625}.  (See Fig.~\ref{fig: stepstone}.)

The reverse-time coalescence of lineages is of central importance in population genetics, particularly in the approach known as coalescent theory~\cite{kingman1982genealogy,wakeley2009coalescent}. 
One of the key estimates of interest in coalescent theory is the expected number of pairwise site differences $\Pi$ between two sampled genomes, which is proportional to the expected time since common ancestry of the two sampled individuals, $T_2$, under the assumption that neutral mutations have accumulated in the (very long) genome at a constant rate since the two lineages diverged. The relation $\Pi \propto T_2$ allows inferences to be made about the population's recent evolutionary past from measured genomic differences in the present, given reliable models of geneaology. The {\it structured} coalescent, which extends coalescent theory to populations with spatial structure (as opposed to well-mixed populations)~\cite{wilkinson1998genealogy}, typically assumes migration rules that produce diffusive dynamics for gene flow. Theoretical studies of the genealogical structure of range expansions have similarly assumed diffusive spatial fluctuations of genetic boundaries (as would be appropriate to a flat front range expansion model; see below) in the interests of analytical tractability~\cite{korolev2010genetic}. Flat front models are equivalent to conventional stepping stone models~\cite{kimura1964stepping} and many exact results are available~\cite{wilkins2002coalescent}.

However, there is strong evidence that  evolutionary dynamics in range expansions are often driven by \textit{superdiffusive} spatial wandering of both genetic sector boundaries and lineages. Hallatschek et al.~\cite{hallatschek2007genetic} measured the mean-square transverse displacement of sector boundaries in \textit{E. coli} growing across hard agar Petri dishes, and found it to scale with the expansion distance $y$ as $y^{2\zeta}$ with wandering exponent $\zeta=0.65\pm 0.05$, greater than the value of $\zeta=1/2$ characterizing diffusive wandering. In both \textit{E. coli} and the yeast species \textit{Saccharomyces cerevisiae}, genetic lineages similarly fluctuate with  wandering exponent $\zeta \approx 2/3$~\cite{ELE:ELE12625}. The same superdiffusive  wandering exponent was found numerically for genetic lineages in an off-lattice model of microbial colony growth~\cite{ELE:ELE12625} and for sector boundaries in a two-species Eden model \cite{korolev2010genetic,saito1995critical}. 
 Consequently, the number of distinct sectors decreases as $y^{-\zeta}$, with $\zeta$ measured to be $\approx 0.67$~\cite{saito1995critical}, a dramatically faster loss of genetic diversity than the $y^{-1/2}$ scaling that would result from diffusive dynamics~\cite{korolev2010genetic}; see Fig.~\ref{fig: stepstone}, where genetically neutral strains are competing.

The underlying cause of this superdiffusive behavior is that the population front profile has a characteristic roughness that increases with time. Because the range expansion causes the front to advance along its local normal direction, stochastically generated protrusions in the front are self-amplifying, and the  lineages and genetic sector boundaries moving with these protrusions experience a faster-than-diffusive average lateral motion. 

Such roughening fronts are characterized by the Kardar-Parisi-Zhang (KPZ) equation~\cite{kpz86,mhkz89}
\begin{equation}
    \partial_t h({\bf x},t)= \nu \nabla^2 h+\lambda (\nabla h)^2/2+\eta({\bf x},t)\,,
\end{equation}
where $h({\bf x},t)$ is the height of the front at position $\mathbf x$ and time $t$, subject to diffusion, growth in the front's local normal direction, and a stochastic noise $\eta({\bf x},t)$. The front roughness $\Delta h \equiv \sqrt{\langle h^2\rangle - \langle h \rangle^2}$ initially grows with time as $t^\beta$, before saturating for a strip of width $L$ as $L^{\beta/\zeta}$.
The scaling exponents, $\beta=1/3$ and $\zeta=2/3$ are known analytically in $d=1+1$ dimensions~\cite{k87,ss10}; this value of the wandering exponent $\zeta$ nicely matches the measured value from experiments and simulations of the microorganism range expansions discussed above.

While there is a wealth of literature on the KPZ equation and its rich universality class~\cite{hhz94,hht15,qs15},  there does not yet exist a similar understanding of the statistics of coalescing space curves -- here, lineages and genetic sector boundaries  -- whose superdiffusive wandering is driven by such KPZ roughening. We term these curves ``KPZ walkers'' in contrast to diffusive random walkers. In developing a quantitative understanding of neutral evolution in a biological range expansion, we are thus led to new questions in statistical physics.

In this work, we numerically investigate the geneological structure of populations with superdiffusive migration of the KPZ walker type, driven by roughening fronts. We are chiefly interested in how the expected time since common ancestry $T_2$ for a pair of individuals depends on spatial separation $\Delta x_0$ at the front, as well as in the probability per unit time $J(\tau|\Delta x_0)$ of lineage coalescence at time $\tau$ in the past, whose first moment $\int_0^{\infty} d\tau\, \tau J(\tau| \Delta x_0) $ equals $T_2(\Delta x_0)$. As a first approach to this problem, our work focuses on neutral evolution from a linear inoculation, avoiding effects such as selection, mutualism/antagonism, and geometrical inflation~\cite{lavrentovich2013radial}, interesting topics of future study.

 We employ a complementary pair of simulation approaches: The first, a lattice-based stepping stone model, introduces front roughness through stochasticity in replication time. In our second approach, we reinterpret the problem of directed paths in random media (DPRM)~\cite{kz87}, a simple and widely-used model from the KPZ unversality class~\cite{kmb91a,kmb91b,hh91}, as a model for range expansions with stochastic variation in organism size.  The DPRM approach can be simulated at  large scales with much less computational expense than our stochastic stepping stone model. We also apply analytical results from the DPRM problem to rationalize the measured asymptotic coalescence behaviors. Finally, we study numerically how environmental heterogeneities temporarily suppress the wandering of KPZ walkers, an effect observed recently in  experiment~\cite{mmn15}.

\section{\label{sec:sim}Methods}

The stepping stone model~\cite{kimura1964stepping} imagines a biological population arranged on a spatial lattice of individually well-mixed subpopulations called ``demes'', each containing $N$ individuals, with exchange of individuals between neighboring demes. We implement the stepping stone model on a triangular lattice with  $N=1$ individual per deme, which models cases in which local fixation of one allele occurs rapidly compared to spatial diffusion~\cite{korolev2010genetic}. 

\begin{figure}
    \centering
    \includegraphics[width=0.9\linewidth]{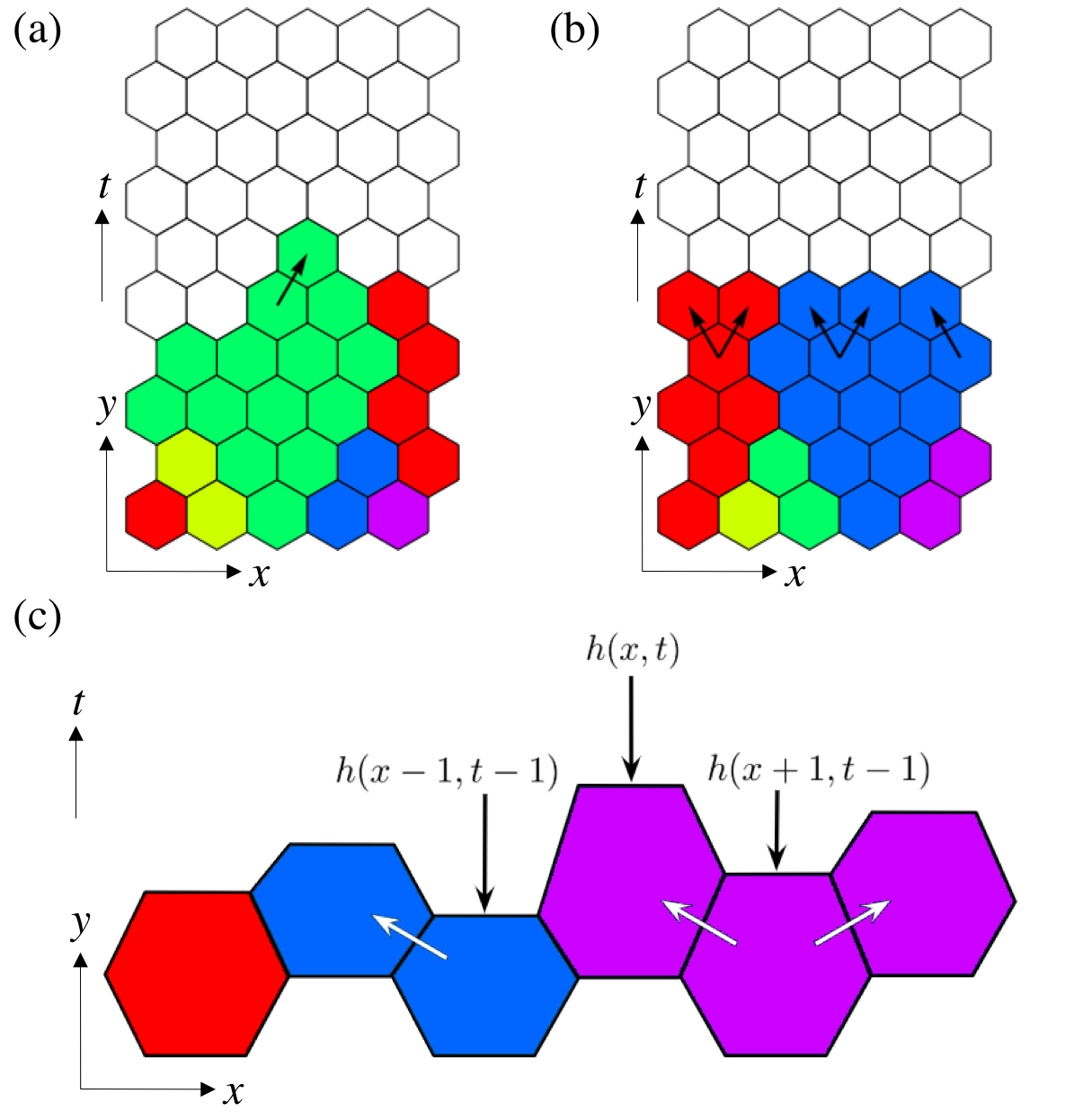}
    \caption{Illlustrations of the the update rules in our numerical models of range expansions. (a,b) The stepping stone model with  deme size $N=1$ on a triangular lattice, using (a) rough front  and (b) flat front update rules. We visualize each individual on the initial line and its descendents with a distinct color. (c) DPRM model of range expansion. At horizontal position $x$, the height of the front in the $y$-direction, $h(x,t)$, is increased by a quantity that depends on the two adjacent heights, namely $ \max\{h(x-t,t-1) + \eta, h(x+1,t-1) + \eta'\}$, where $\eta$, $\eta'$ are zero-mean stochastic noise terms that cause front roughness. The nearest neighbor cell which maximizes the above relation is chosen to reproduce, and passes on its allele label (denoted by the color), as represented by white arrows in the illustration.}
    \label{fig: rules_of_evolution}
\end{figure}

\begin{figure}
\centering
\includegraphics[width=0.9\linewidth]{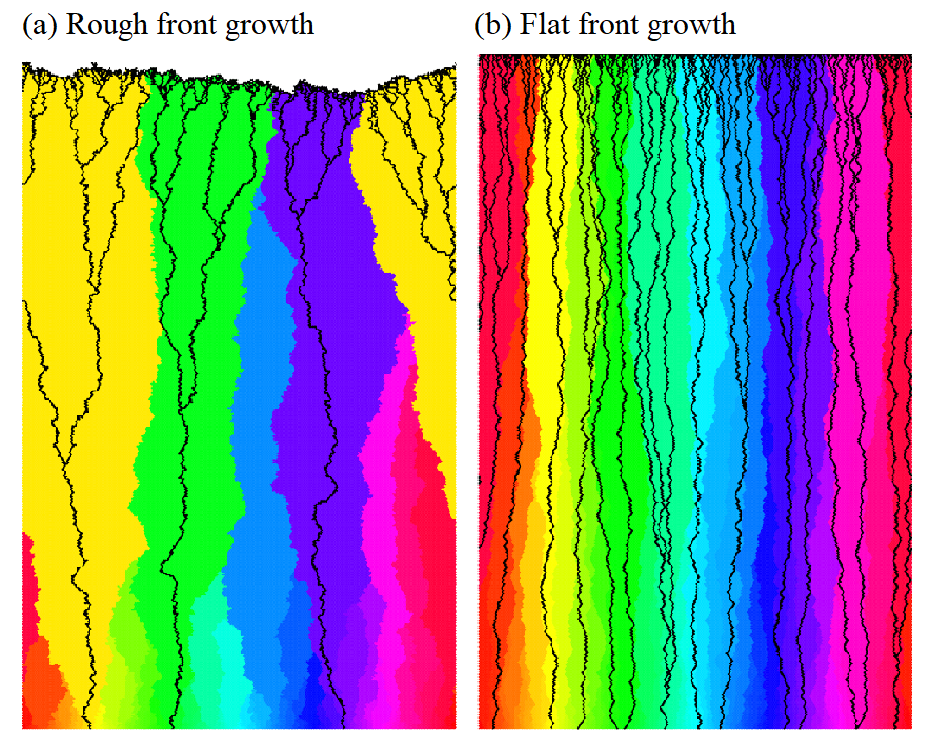}
\caption{
Range expansions generated by the stepping stone model, using the (a) rough front and (b) flat front  update rules, with periodic boundary conditions in the horizontal direction. The colors represent allele labels, while the black lines mark the genetic lineages. Time runs upward in both cases. Note that there are fewer {\it sectors} at the top (genetic coarsening), but fewer {\it lineages} at the bottom (lineage coalescence). {Typical coalescence rates are much larger in (a) than in (b).} 
}
\label{fig: stepstone}
\end{figure}

As an initial condition, we take the lattice of demes in two dimensions to be unpopulated except for a linear inoculation ``homeland''. Once a deme is populated, its allele remains unchanged thereafter, as in the microbial experiments on agar plates, where cell divisions occur only near the frontier, so that the spatial pattern of alleles is effectively frozen behind the front~\cite{hallatschek2007genetic}. 
We choose as our update rule that of the Eden model~\cite{eden1961two} for two-dimensional growth processes: One site is chosen at random from among all  occupied sites with some empty neighbor site, and the allele is copied from the chosen occupied site into a randomly chosen empty neighbor (Fig.~\ref{fig: rules_of_evolution}a)~\cite{endnote1}.  By introducing stochasticity in the replication time, this procedure generates an irregular interface between the occupied and empty regions (see Fig.~\ref{fig: stepstone}a), simulating a rough front range expansion. By contrast, the expansion front remains flat if the update rule fills an entire row in parallel (Fig.~\ref{fig: rules_of_evolution}b), with each newly filled site inheriting the allele marker of one of its two filled neigbhors below, chosen randomly with equal probability. The dynamics in Fig.~\ref{fig: rules_of_evolution}b is equivalent to a one-dimensional stepping stone model in discrete time with deme size $N=1$.

The second model, DPRM~\cite{kz87}, arises from the problem of finding a minimal-energy directed path through a random energy landscape $\eta(x,t)$. Directed paths must propagate in the `time' direction $t$, but can fluctuate in the spatial direction $x$.  
We can reinterpret DPRM as an alternative model of range expansions with roughening fronts. 
In Fig.~\ref{fig: rules_of_evolution}c, we illustrate that the accumulated ``energy'' of the directed path, characterized by the KPZ equation, can be mapped to the height of a range expansion front. In this mapping, the stochastic noise $\eta$ corresponds to fluctuations in the lengths of individual microbes in the direction of average propagation $y$, about a mean length $\ell$. 
An allele label is added to each site, as in the stepping stone model. The height of the front $h(x,t)$ is updated according to 
\begin{equation}
    h(x,t) = \ell\,+\, \max\{h(x-t,t-1) + \eta, h(x+1,t-1) + \eta'\},
    \label{eq: tm}
\end{equation}
where $\eta,$ $\eta'$ are zero-mean, independent and identically distributed  random variables. Each site at time $t$ is then filled by the offspring of one of its nearest neighbours from time $t-1$, and inherits the corresponding allele label. The choice of competing mother cells is taken to be the cell that optimizes the relation in Eq.~\ref{eq: tm}. 

\begin{figure}
\includegraphics[width=0.9\linewidth]{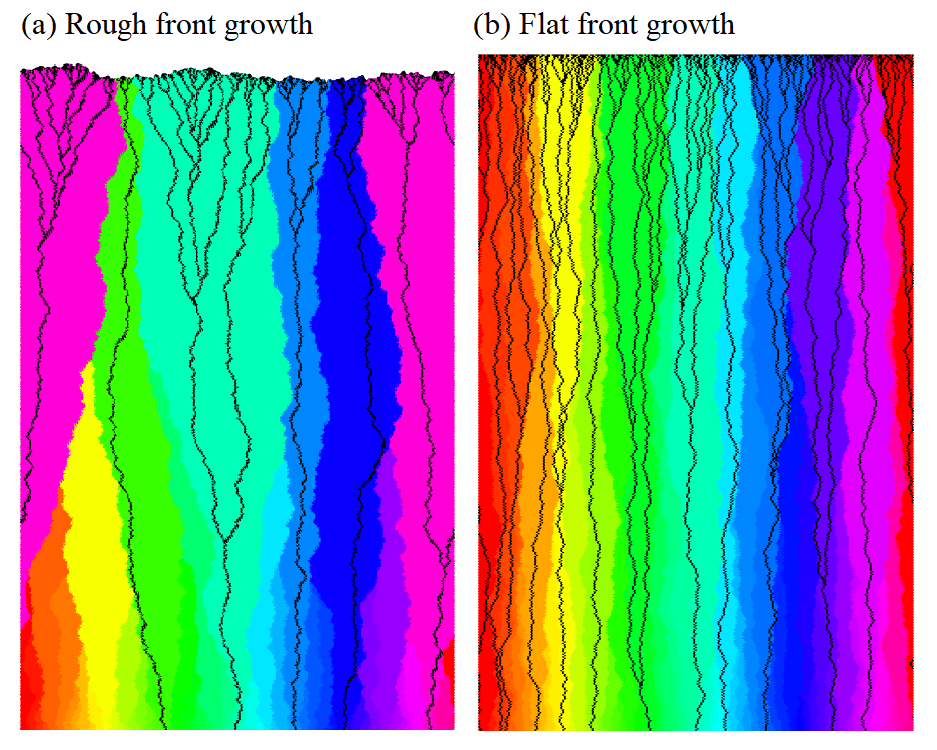}
\caption{Range expansions generated by the DPRM model, with periodic boundary conditions in the horizontal direction, as in Fig.~\ref{fig: stepstone}. The colors represent allele labels, while the black lines mark the genetic lineages. In contrast to the flat front case (b), the rough front case (a) with the same number of generations shows a significantly faster decrease in genetic diversity, and {much larger coalescence rates}, similar to Fig.~\ref{fig: stepstone}. 
}
\label{fig: rf_dprm}
\end{figure}

Thus, while replication time is constant in this model,  front roughness is generated by stochasticity in cell size, with larger size favored for propagation. While we assume that the mean cell size at time of division for the microbe in question has already evolved to a fitness maximum, variance in the cell size leads to front roughness and accelerated loss of genetic diversity (Fig.~\ref{fig: rf_dprm}a).

Note that if we fix $\eta$ to have zero variance, and instead choose the mother cell at random between the left- and right-neighbours, we recover a flat front range expansion with diffusive dynamics associated with lineages and genetic boundaries (Fig.~\ref{fig: rf_dprm}b). Also, if we reduce the system width to a single organism, the front height $h(x,t)$ performs a random walk about the determnistic value $\ell t$, the variance growing linearly in $t$ with slope given by the variance in $\eta$. A dramatic experimental realization of such a scenario in \textit{E.\ coli} was demonstrated by the ``mother machine'' of Wang \textit{et al.} \cite{wang2010robust}: Bacteria growing and dividing in narrow channels, quasi-one-dimensionally, show a range of cell sizes, with the overall growth rate following a Gaussian distribution.

In both the rough front stepping stone model and the DPRM model, lineages and sector boundaries have superdiffusive lateral fluctuations with wandering exponent $\zeta=2/3$~\cite{k87,kz87,ss10,korolev2010genetic,saito1995critical}. For DPRM models, this behavior is well-known as the transverse fluctuations of the minimal-energy directed path. In contrast, for the flat front stepping stone model and the zero-noise limit of DPRM, the lateral fluctuations of lineages and sector boundaries are merely diffusive, $\zeta=1/2$. 

This superdiffusive behavior has stark consequences for the genetic structure of the population. Comparing the flat front and rough front realizations for the stepping stone model in Fig.~\ref{fig: stepstone} and for the DPRM model in Fig.~\ref{fig: rf_dprm}, we see striking differences in both the coalescing lineage trees and  the decay in the number of surviving monoclonal sectors. Genetic diversity is lost much more rapidly in the rough front case, and nearby individuals at the front are much more likely to have a common ancestor in the recent past, reflecting {much larger coalescence rates}. 


Further details about the numerical implementation of these two methods are given in the Supporting Information.

\section{\label{sec:results}Results and Discussion}

\subsection{\label{sec:coal}Coalescence of lineages}

\subsubsection{Rate of coalescence $J(\tau|\Delta x_0)$}

For two lineages separated by $\Delta x_0$ at the front, $J(\tau|\Delta x_0)$ is the probability per unit time for them to coalesce in a common ancestor at reverse time $\tau$. In the diffusive case, on an infinite line, this is the well-known coalescence rate for two diffusive random walkers with diffusion constant $D$~\cite{redner2001guide}:
\begin{align}
J_{\mathrm{diff}} (\tau|\Delta x_0) &= \frac{1}{\sqrt{8\pi D}} \frac{1}{\tau} \left(\frac{\Delta x_0^2}{\tau}\right)^{1/2}  \exp{\left[-\frac{1}{8 D} \left(\frac{\Delta x_0^2}{\tau}\right) \right] }. \label{eq: Jdiff}
\end{align}
Results such as Eq.~\ref{eq: Jdiff}, valid here for flat front models, will serve as a useful guide to our
investigations of more complex coalescent phenomena at rough frontiers. In population
genetics, systems analogous to our flat front models also arise in the continuum limit of one-dimensional Kimura-Weiss stepping stone models~\cite{kimura1964stepping}. As reviewed in Ref.~\cite{korolev2010genetic}, many exact
results for quantities such as the heterozygosity correlation function and coalescent times are
available~\cite{bde02, m75, n74, wh98}. The $x$-coordinate of stepping stone models represents the horizontal
axis of flat front simulations such as those displayed in Fig.~\ref{fig: stepstone}b and~\ref{fig: rf_dprm}b, while its time
coordinate maps on to the $y$-axis. Nullmeier and Hallatschek have used a stepping stone model
to study how coalescent times change in 1-dimensional 
populations when one boundary of a habitable domain
moves in a linear fashion due to, say, a changing climate~\cite{nh13}. Results from this later
investigation could thus be reinterpreted as applicable to a two-dimensional range expansion in a
trapezoidal domain, in the flat front approximation with diffusive genetic boundaries.

For superdiffusive lineages, however, the full expression for $J(\tau|\Delta x_0)$ is not known.
We focus instead on its asymptotic behaviors using predictions from DPRM and intuition gained from the diffusive case. For lattice models like those in Fig.~\ref{fig: rules_of_evolution}, it will be convenient to measure distances $\Delta x_0$ in units of the space-like direction $x$, and $\tau$ in units of the fundamental step in the time-like direction, which amounts to scaling out the analog of the diffusion constant in Eq.~\ref{eq: Jdiff}. We expect on theoretical grounds that $J$  depends on $\Delta x_0$ only through the combination $\Delta x_0/\tau^\zeta$, with exponent $\zeta=2/3$ as opposed to $\zeta=1/2$ in the diffusive case.

\begin{figure}
\includegraphics[width=\linewidth]{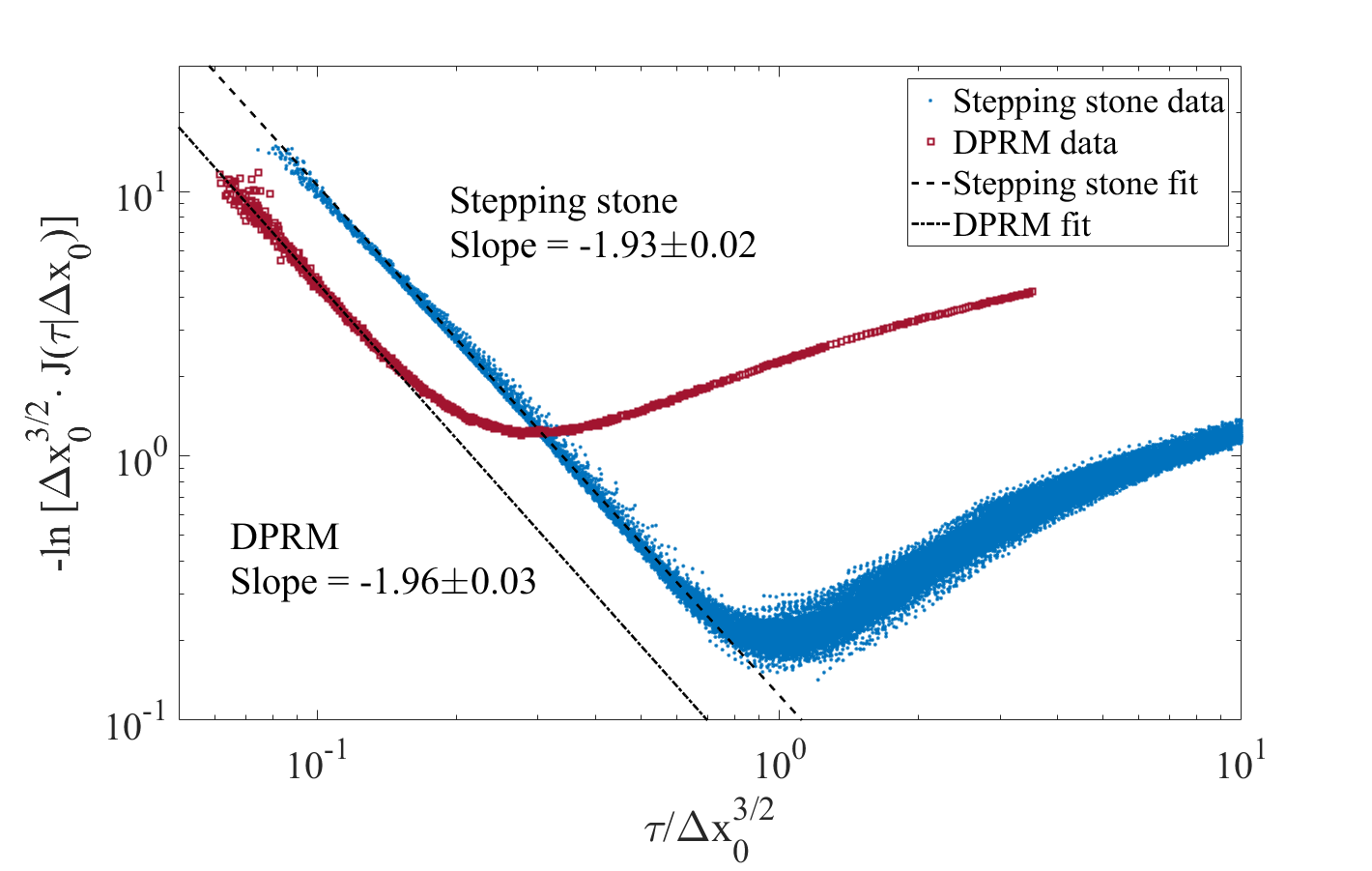}
\caption{Log-log plot of $-\ln[\Delta x_0^{3/2}J(\tau|\Delta x_0)]$ vs.\ the KPZ-rescaled variable $\tau/\Delta x_0^{3/2}$ for lineages in the stepping stone model and for DPRM. {Here, we focus on the regime $\Delta x_0 \ll L$, to avoid finite size effects associated with periodic boundary conditions.} 
Asymptotically for $\tau/\Delta x_0^{3/2} \ll 1$, the relationship is linear, indicating an exponential form for $J(\tau|x_0)$. The fitted slopes are $-1.93\pm0.02$ for stepping stone, and $-1.96\pm0.03$ for DPRM, providing measurements of $\gamma$ as defined in Eq.~\ref{eq: J_exponentialform}. (For comparison, the DPRM theory predicts a slope of $-2$.) 
}
\label{fig: exp_fit}
\end{figure}

First, we consider the regime $\tau/\Delta x_0^{3/2} \ll 1$, representing
rare coalescence events where lineages located far apart at the front can be traced back to a recent common ancestor.
For the analogous  regime of $\tau/\Delta x_0^2 \ll 1$ in the diffusive case, the coalescence rate
behaves as 
$J_{\mathrm{diff}}(\tau|\Delta x_0) \sim \exp[-(\Delta x_0/\tau^{1/2})^2]$.
We  hypothesize a similar decay for the superdiffusive case, as 
\begin{equation}
    J(\tau|\Delta x_0) \sim \exp\left(-\left(\frac{\Delta x_0}{\tau^{2/3}}\right)^{\gamma'}\right)=\exp\left(-\left(\frac{\tau}{\Delta x_0^{3/2}}\right)^{\gamma}\right) 
    \label{eq: J_exponentialform}
\end{equation}
for some exponent $\gamma=-\frac{2}{3} \gamma'$. In Fig.~\ref{fig: exp_fit}, we plot $-\ln [\Delta x_0^{3/2} J(\tau|\Delta x_0)]$ vs.\ $\tau/\Delta x_0^{3/2}$ for both the stepping stone model and DPRM on a log-log scale, so that Eq.~\ref{eq: J_exponentialform} predicts a linear plot with slope $\gamma$. At small $\tau/\Delta x_0^{3/2}$, both sets of data appear linear, confirming the above hypothesized  form. 
The slopes in the linear regime provide estimates of 
$\gamma=-1.96\pm0.03$ for DPRM and $-1.93\pm0.02$ for the stepping stone model.

In fact, we can analytically derive this exponential form, including the value of $\gamma$, using the 
known distribution of directed path endpoints in DPRM~\cite{fqr13}. The  calculation, given in the Supporting Information, shows that
\begin{align}
J(\tau|\Delta x_0) \sim \frac{1}{\tau} \left( \frac{\Delta x_0}{\tau^{2/3}} \right)^{1/2} \exp \left( - \frac{c}{4} \left(\frac{\Delta x_0}{\tau^{2/3}}\right)^{3} \right),    \label{eq: J_saddlepoint}
\end{align} 
where $c$ is a constant of order unity.
For $\tau/\Delta x_0^{3/2} \ll 1$, the leading asymptotic behavior of $J(\tau|\Delta x_0) \sim \exp(-\frac{1}{4} c(\Delta x_0/\tau^{2/3})^3)$  thus corresponds to $\gamma'=3$, $\gamma=-2$. From the numerical results in Fig.~\ref{fig: exp_fit}, we see from DPRM that $\gamma \approx -1.96\pm 0.03$, and from the rough front stepping stone model we compute $\gamma \approx -1.93\pm 0.02$. Both numerical results are in good agreement with the analytically derived prediction.


In the opposite regime of $\tau/\Delta x_0^{3/2} \gg 1$, we can again hypothesize a form for $J$ in analogy with the diffusive case, for which Eq.~\ref{eq: Jdiff} shows $J_{\mathrm{diff}}(\tau|\Delta x_0) \sim \tau^{-1} (\Delta x_0/\tau^{1/2})$. For KPZ walkers, the analogous form is 
\begin{equation}\label{eq:smallDx}
J(\tau|\Delta x_0) \sim \frac{1}{\tau} \left( \frac{\Delta x_0}{\tau^{2/3}} \right)^{\alpha'} = \frac{1}{\Delta x_0^{3/2}} \left( \frac{\tau}{\Delta x_0^{3/2}} \right)^{\alpha}, 
\end{equation}
for some exponent $\alpha=-(1+\frac{2}{3} \alpha')$. 
Although the 
expression in Eq.~\ref{eq: J_saddlepoint} is consistent with this form, that result is obtained by assuming the two KPZ walkers to be independent (valid at small $\tau/\Delta x_0^{3/2}$), so there is no reason to expect 
the apparent value of $\alpha'=1/2$, $\alpha=-4/3$ to hold for $\tau/\Delta x_0^{3/2} \gg 1$.

\begin{figure}
\includegraphics[width=\linewidth]{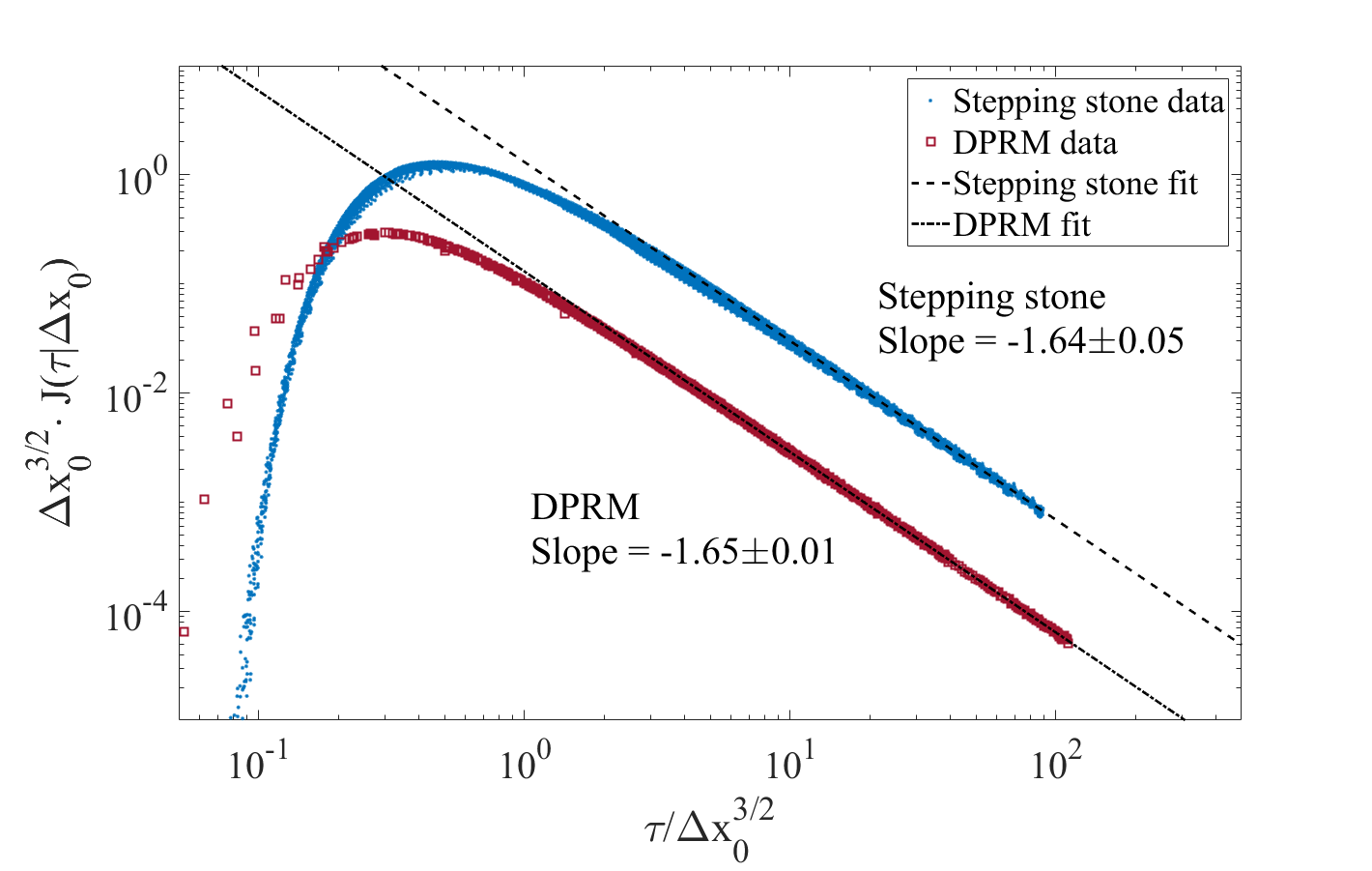}
\caption{Log-log plot of $\Delta x_0^{3/2} J(\tau|\Delta x_0)$ vs. the KPZ-rescaled variable $\tau/\Delta x_0^{3/2}$ for lineages in the stepping stone model and for DPRM. For $\tau/\Delta x_0^{3/2} \gg 1$, the exponent of the power-law decay (Eq.~\ref{eq:smallDx}) is extracted from a linear fit to the numerical data, yielding $\alpha = -1.64\pm0.05$ for stepping stone, and $\alpha =- 1.65\pm0.01$ for DPRM. {As in Fig.~\ref{fig: exp_fit}, we work in the limit $\Delta x_0 \ll L$ to avoid effects due to periodic boundary conditions.} 
}
\label{fig: pl_fit}
\end{figure}

The rate of coalescence for the two computational approaches in this regime is plotted in Fig.~\ref{fig: pl_fit}. The asymptotic behavior is consistent with the hypothesized power-law decay. The exponent $\alpha$ is determined numerically to be $\alpha = -1.64\pm0.05$ for the stepping stone model, and $\alpha = - 1.65\pm0.01$ for DPRM, giving good agreement between the two models. Furthermore, these values do not rule out the possibility that $\alpha = - 5/3$, $\alpha'=1$, which would give the noteworthy conclusion that $J(\tau|\Delta x_0)$ is linear in the separation $\Delta x_0$, just as in the diffusive case.

\subsubsection{Expected time to coalescence $T_2$}
For a range expansion that has proceeded for a time $t_{\max}$ after a linear inoculation, if two lineages separated by $\Delta x_0$ share a common ancestor on the initial line, we can calculate their expected time to coalescence (time since common ancestry) as
\begin{equation}
    T_2 (\Delta x_0, t_{\max}) \equiv \frac{\int_0^{t_{\max}} d\tau \; \tau J(\tau|\Delta x_0)}{\int_0^{t_{\max}} d\tau \; J(\tau|\Delta x_0)}.
\end{equation}
Note that the denominator represents normalization by the probability that the two lineages do indeed coalesce.

\begin{figure}
\includegraphics[width=\linewidth]{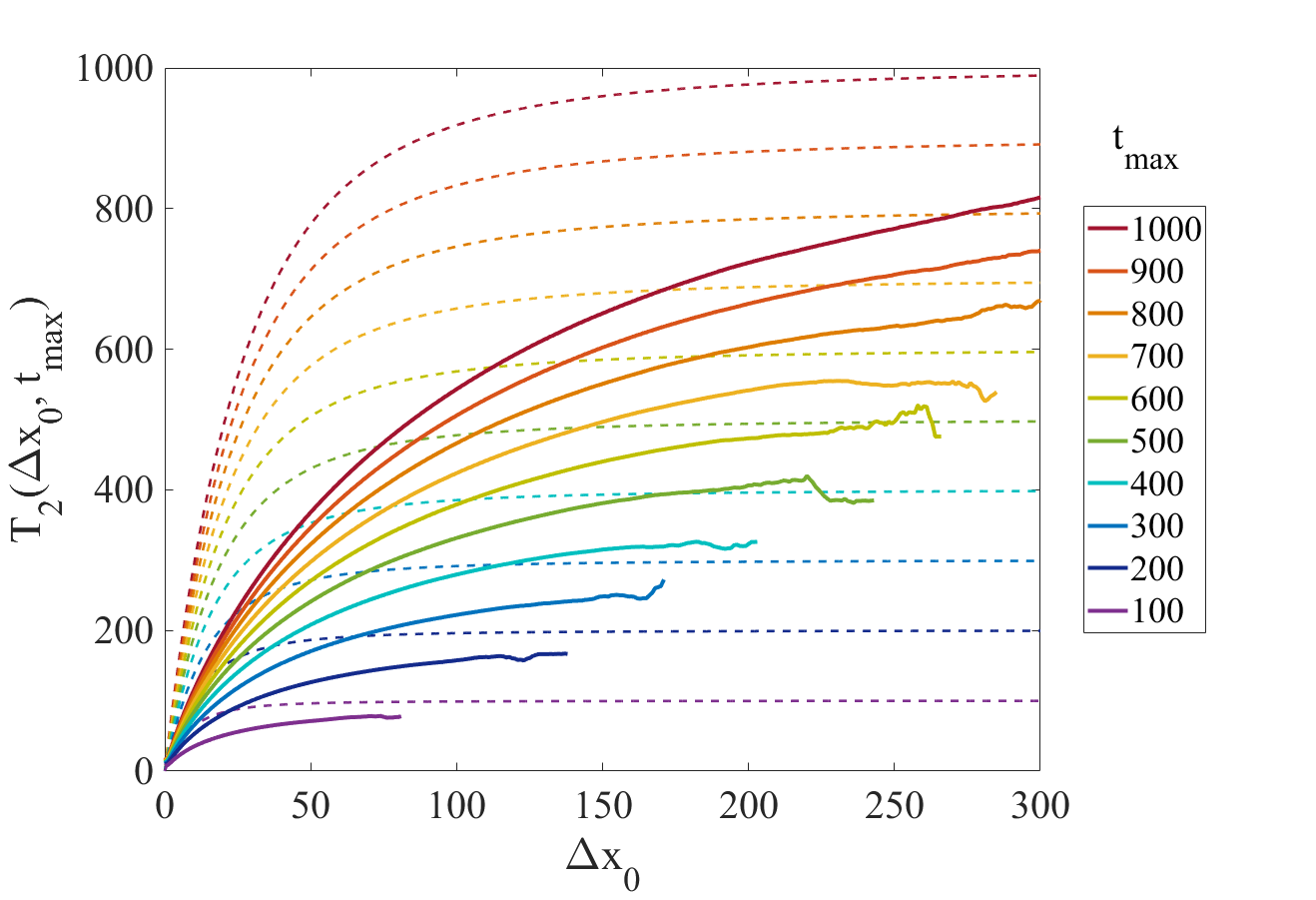}
\caption{Average time $T_2$ since common ancestry for pairs of individuals with separation $\Delta x_0 \ll L$ at the front and some common ancestor in the past, for a range of system expansion times $t_{\max}$. Solid lines represent numerical data for KPZ walkers in the stepping stone model
, and dashed lines represent analytical predictions for diffusive walkers with the same parameters. The plateau values are simply $t_{\text{max}}$.}
\label{fig: T2}
\end{figure}

In the case of diffusive lineages, Eq.~\ref{eq: Jdiff} leads to an analytic expression for $T_2$,
\begin{equation}\label{eq: T2_diff}
   \frac{ T_{2,{\text{diff}}}(\Delta x_0, t_{\max})}{t_{\max}}=\left(\frac{\Delta x_0^2}{8Dt_{\max}}\right) \frac{\Gamma \left[ -1/2, \Delta x_0^2/8Dt_{\max} \right]}{\Gamma \left[ 1/2, \Delta x_0^2/8Dt_{\max} \right]},
\end{equation}
where $\Gamma(x,y)$ is the incomplete gamma function. In Fig.~\ref{fig: T2} we compare the numerical $T_2$ data for KPZ walkers in the rough front stepping stone model 
with the analytical prediction from the diffusive case under the same conditions. While for large $\Delta x_0$, $T_2$ approaches $t_{\max}$, the behavior for
small $\Delta x_0$ is controlled by the scaling in Eq.~\ref{eq:smallDx}: an approximately linear scaling
leading to $T_2\sim \Delta x_0 t_{\max}^{1-\zeta}$.
We see that lineages with the same separation $\Delta x_0$ coalesce much faster on average when they behave as KPZ walkers, and that this difference becomes more pronounced for large $t_{\max}$, as is evident qualitatively from Figs.~\ref{fig: stepstone} and~\ref{fig: rf_dprm}.
The scaling of $T_2$ for KPZ walkers can be written in a form analogous to Eq.~\ref{eq: T2_diff}, and reflects the KPZ transverse scalings inherent in the system (see Supporting Information).

In biological terms, common ancestry is expected to be more recent under rough front dynamics than under diffusive dynamics. As a result, assuming a constant rate of neutral mutations, the number of differences {$\Pi(\Delta x_0)$} 
between pairs of two sampled genomes at the front is expected to increase more slowly with separation $\Delta x_0$ along the front. This anomaly arises because we expect the habitat to be populated by the offspring of a small number of common ancestors, which decays as $t^{-2/3}$ for KPZ walkers, rather than the $t^{-1/2}$ decay characterizing diffusive random walkers, where $t$ is the time since the initial inoculation.

\subsection{\label{sec:obs}Environmental Heterogeneities}
The presence of environmental heterogeneities in the habitat can have a significant impact on a range expansion, including on the front shape and propagation speed, and on the genetic diversity at the front.
A prototypical example of environmental heterogeneity is the obstacle, a nutrient-depleted zone, that the population must grow around rather than through. Range expansions around an obstacle were studied experimentally and via simple geometrical optics ideas by M\"obius et al.~\cite{mmn15} (see also~\cite{tesser2016population}).  
A notable feature of  the experimental (and numerical)  results from Ref.~\cite{mmn15} is that the sector boundary which forms at the apex of the obstacle shows suppressed transverse fluctuations compared to all other sector boundaries. As the front propagates past the obstacle, a component of its velocity is directed inward from both sides. This in effect pins the sector boundary to the middle, at a kink in the front, and suppresses this sector boundary's fluctuations. 

\begin{figure}
\begin{tabular}{m{0.5\linewidth} m{0.5\linewidth}}
\centering \includegraphics[height=1.2 \linewidth]{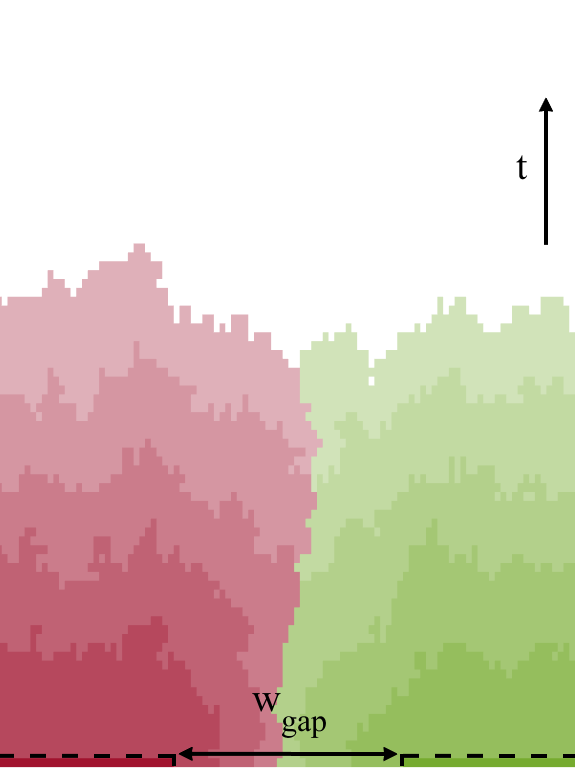} &   \includegraphics[height=1.2 \linewidth]{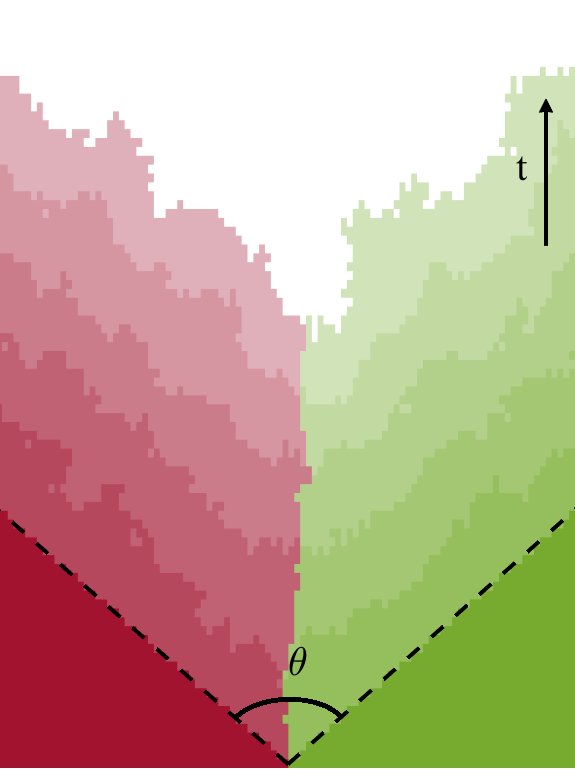}\\ 
\centering (a) & \centering (b) 
\end{tabular}
\caption{Geometries of the sector boundary between two alleles (labeled red and green). The initial inoculations are marked by dashed lines. (a) Illustration of the gap geometry: A segment of width $w_{\mathrm{gap}}$ is left unpopulated initially, separating the two alleles which grow from an otherwise flat initial condition. The width $w_{\text{gap}}$ could represent, say, the width of a square obstacle that terminates at time $t=0$, or the size of an interval along the horizontal $x$-direction where all organisms are removed by an environmental trauma. (b) Illustration of the wedge geometry: The initial population occupies two triangular regions whose growth fronts meet at a wedge angle $\theta$. In both systems, the two alleles meet at a single sector boundary, along which fluctuations are suppressed. The front of the range expansion is illustrated for a series of equally spaced time values $t$, with lighter shades representing later times.}
\label{fig: gapwedge}
\end{figure}

\begin{figure}
\includegraphics[width=\linewidth]{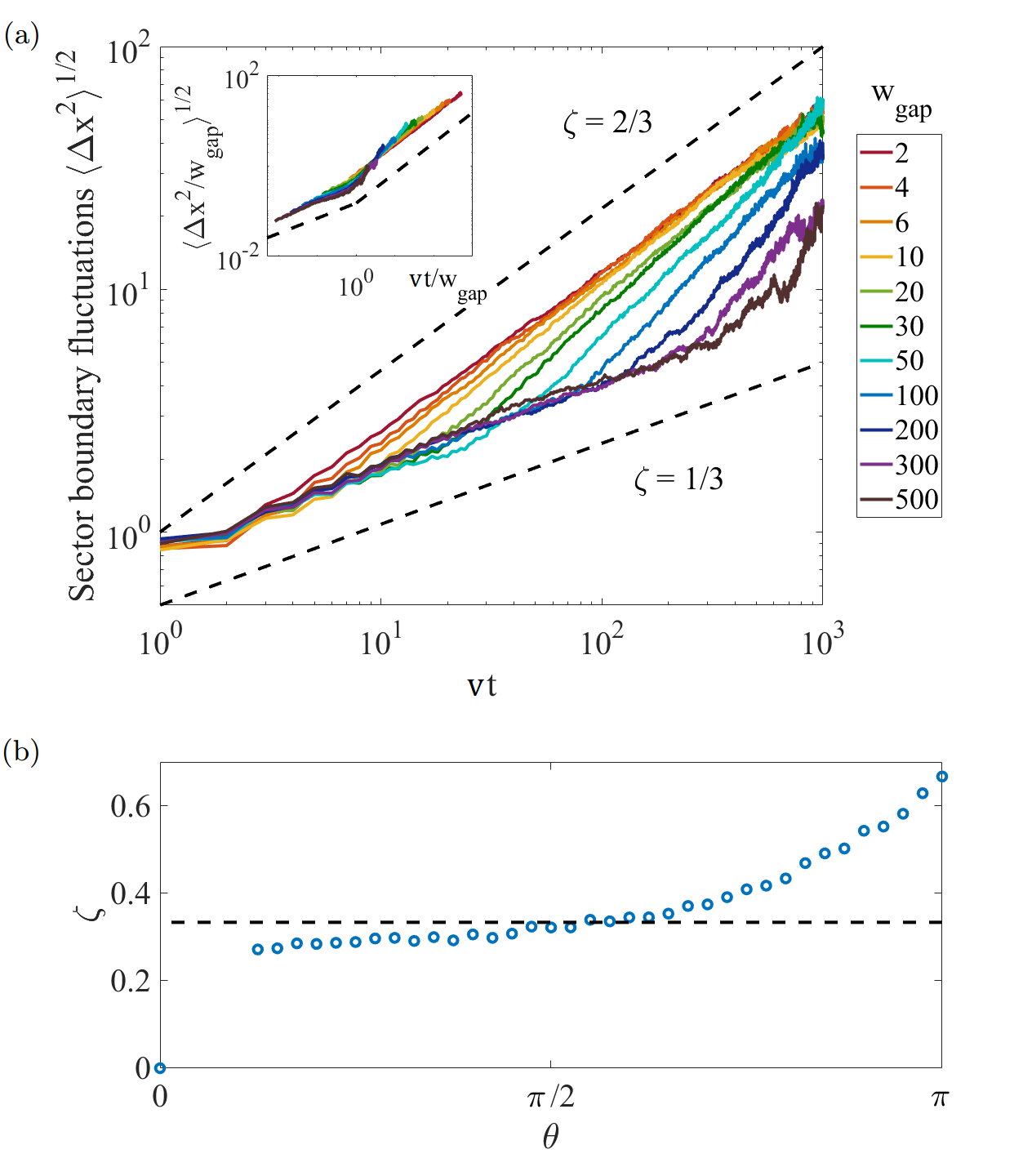}
\caption{
(a) Log-log plot of fluctuations of the sector boundary $\langle \Delta x^2 \rangle^{1/2}$ vs. vertical distance along the sector boundary $vt$ in the gap geometry for a range of gap sizes $w_{\text{gap}}$. 
Fits to a power law scaling form $\langle \Delta x^2 \rangle^{1/2} \sim t^\zeta$ yield exponents varying from $\zeta \approx 1/3$ to $\zeta \approx 2/3$, with a crossover region in between.
Inset: Data collapse after rescaling with respect to $w_{\text{gap}}$. By geometrical arguments, $vt/w_{\text{gap}}$, where $v$ is the average front speed, is a measure of the angle of incidence of the fronts as determined by a constant speed or ``geometrical optics'' model. We see a reasonably good collapse across many different gap sizes, with $\zeta \approx 1/3$ for $vt/w_{\text{gap}} < 1$, and $\zeta \approx 2/3$ for $vt/w_{\text{gap}} > 1$.
(b) Wandering exponent $\zeta$ as a function of the angle of incidence $\theta$ in the wedge geometry. As $\theta$ increases from 0 to $\pi$, the wandering exponent increases smoothly from approximately $\zeta = 1/3$ (marked by the dashed line) to the KPZ value of $\zeta = 2/3$.}
\label{fig: obstacles}
\end{figure}

Here, we study these suppressed fluctuations in greater detail using the  stepping stone model with a rough front. A gap of width $w_{\mathrm{gap}}$ of unoccupied sites is left in the initially populated line, providing a simplified representation of a range expansion past an obstacle of such width, or the result of an environmental trauma (Fig.~\ref{fig: gapwedge}a). By considering only two ``alleles'' (colors), we can track the wandering of the single sector boundary that forms approximately above the center of the obstacle. As shown in Fig.~\ref{fig: obstacles}a, the effective wandering exponent $\zeta$ is suppressed from the usual value of $2/3$, to $\zeta \approx 1/3$ for times $vt\lesssim w_{\mathrm{gap}}$, where $v$ is the average front velocity. At later times, as the kink in the front heals and the average front normals return to the vertical, $\zeta$ recovers the expected value of $2/3$ for KPZ genetic boundaries. 
Notably, the effective $\zeta$ appears to exceed $2/3$ in an intermediate transitory regime when $vt\approx w_{\mathrm{gap}}$.

To gain further insight into this changing wandering exponent, we modify the numerical experiment to a wedge geometry (Fig.~\ref{fig: gapwedge}b). 
This allows us to fix the kink angle $\theta$ to be a constant value, as opposed to the gap geometry where the kink  heals from some initial $\theta_0$ toward $\pi$ with increasing time.
Now, the stepping stone model with deme size of 1 is, in essence, identical to the Eden model on a triangular lattice, 
with the added complication of tracking different genotypes.
The boundary between two Eden clusters meeting at an angle $\theta$ has previously been studied,~\cite{dd91}. 
The transverse fluctuations scale as $t^\zeta$, where $t$ is the simulation time, and the wandering exponent 
$\zeta$ was conjectured to be
\begin{equation}
    \zeta(\theta) = \left\{ \begin{array}{ll}
    1/3,  & \theta < \pi, \\
    2/3,  & \theta = \pi, \\
    1,  & \theta > \pi.
    \end{array} \right.
    \label{eq: zeta}
\end{equation}
The value $\theta = \pi$ corresponds to two Eden clusters growing side by side with flat initial conditions, 
in which case one recovers the KPZ value of $\zeta = 2/3$ as expected.

The regime $\theta < \pi$ is of relevance to range expansions with obstacles.
Heuristically, the sector boundary becomes pinned by the two Eden clusters growing into each other, and the usual KPZ transverse fluctuations are suppressed.
Instead, the fluctuations which dominate are those of the propagating fronts themselves, which scale with the KPZ growth exponent $\beta = 1/3$ rather than the wandering exponent $\zeta = 2/3$. 

The original simulations which led to the estimates in Eq.~\ref{eq: zeta} sampled only 3 points in the range $\theta < \pi$, namely $\theta = \pi/3$, $\pi/2$, and $2\pi/3$~\cite{dd91}.
We expand on this previous work by fitting to an an effective $\zeta(\theta)$ for many more values of $\theta$.
The results plotted in Fig.~\ref{fig: obstacles}b indicate a smooth crossover between 
$\zeta = 1/3$ and $\zeta = 2/3$ as 
$\theta$ increases from 0 to $\pi$. A heuristic explanation for this change in $\zeta$ is given in the Supporting Information. 
The results from the wedge geometry are qualitatively consistent with the $\zeta$ values measured from the ``gap geometry.''
As the range expansion propagates around an obstacle, the fronts from either side meet at some angle $\theta_0 < \pi$, which can be predicted by a deterministic model of constant-speed propagation for wavefronts in the same geometry, inspired by geometrical optics~\cite{mmn15}.
The incident angle increases up to $\theta = \pi$ as the kink in the front heals.
Therefore, for the sector boundary formed after the obstacle, we expect the wandering exponent to initially take some value $\zeta < 2/3$, and then slowly recover to $\zeta = 2/3$.
The kink has healed when the fluctuations of the front (perpendicular to the direction of propagation) are comparable to the size of the dip.


\section{\label{sec:conc}Conclusion and Outlook}

The propagating front of a range expansion is expected to roughen over time, and in this work we have connected the population genetics of such range expansions with new calculations in statistical physics models from the KPZ universality class.  We have shown, through both DPRM calculations and a stepping stone model with rough fronts, that  the superdiffusive ``KPZ walkers'' describing genetic lineages have coalescence statistics whose limiting behaviors are qualitatively, but not at all quantitatively, similar to those of coalescing diffusive random walkers. In the limit of large separation or small time in the past, the coalescence rate for KPZ walkers decays as $J\sim \exp[-(\tau/\Delta x_0^{3/2})^{-2}]$, in contrast to the scaling $J_{\mathrm{diff}}\sim \exp[-(\tau/\Delta x_0^2)^{-1}]$ for the diffusive case in the same limit. 

In the opposite limit of small separation or large time in the past, we find that $J$ varies algebraically as $\tau^{-1} ( \Delta x_0/\tau^{2/3})^{\alpha'}$ with $\alpha' \approx 1$, whereas diffusive random walkers coalesce according to the form $J_{\mathrm{diff}}\sim \tau^{-1} (\Delta x_0/\tau^{1/2})$.

From these numerically measured coalescence rates, we have calculated the expected time $T_2$ since common ancestry for pairs of individuals as a function of their spatial separation, an important quantity in population genetics. The superdiffusive wandering of lineages suppresses $T_2$ significantly compared to estimates based on diffusive dynamics. Our results go beyond the known scaling difference between diffusive and KPZ lineages and genetic boundaries, and provide quantitative information about how front roughness leads to more recent, and fewer, common ancestors for the ``pioneers'' comprising the front. 

We have also used the stepping stone model to explain how environmental heterogeneities can alter this superdiffusive dynamics, even leading to time regimes with subdiffusive dynamics. Our results explain the suppressed fluctuations of genetic sector boundaries behind an obstacle observed in recent experimental work, and connect them with prior numerical work on Eden model growth. The effect of obstacles can be viewed as a competition between the usual roughening of the front, which favors the KPZ wandering exponent $\zeta=2/3$, and the collision of two segments of the front propagating around either side of the obstacle, which suppresses $\zeta$ toward the value of the front roughness exponent $\beta=1/3$. 

Going forward, our calculations of $J$ and $T_2$ for KPZ walkers in a totally uniform environment will be valuable as a standard against which deviations can be measured, to reveal the effects of various realistic complications. These complications include end effects from habitat boundaries~\cite{nh13, wilkins2002coalescent}, selectively advantageous or deleterious mutations, mutualism or antagonism between subpopulations~\cite{lavrentovich2014asymmetric}, geometrical inflationary effects in radial expansions \cite{lavrentovich2013radial}, and more complex heterogeneities in the environment~\cite{mmn15}. 

On the latter topic, we have made headway here by studying a simplified representation of an obstacle as a prototypical environmental heterogeneity, which already illustrates the subtle issue of locally suppressed fluctuations. It will be interesting to extend this analysis of Eden model growth to situations with multiple obstacles, and with other types of heterogeneities such as nutrient ``hotspots''~\cite{tesser2016population} and uneven topography~\cite{beller2018evolution}. The dynamics can also be made more sophisticated by increasing the number of organisms per deme above $N=1$, and reintroducing aspects of the original stepping stone model's migration dynamics between neighboring demes~\cite{kimura1964stepping}.  

From the perspective of statistical physics, range expansions provide
not only an experimental testing ground for the predictions of KPZ scaling, but also an
incentive to introduce and explore variants of rough growth. For example, the coalescing domain boundaries in Fig.~1 qualitatively resemble coarsening of
domains in a multi-component growth process~\cite{k99}, and should be
quantitatively described by the coupling of directed percolation (of genetic domains) to
the rough interface~\cite{hk18}.

Finally, our results have drawn upon connections between two quite different processes in the KPZ universality class, the rough front stepping stone model and DPRM, to obtain quantitative insights about biological experiments that can be realized in the laboratory. We hope that this work will inspire future investigations to seek other useful links between disparate model systems that shed light on the evolutionary dynamics of rough front range expansions, a problem with much fertile territory.  

\acknowledgments{
DRN and DAB acknowledge frequent conversations with W. M\"obius during the early stages of this investigation. MK and SC acknowledge support from NSF through grant DMR-1708280. Work by DRN and DAB was supported in part by the NSF through Grant DMR-1608501 and by the Harvard Materials Science Research and Engineering Center through Grant DMR-1435999. This research was initiated during a visit to the Kavli Institute for Theoretical Physics supported through Grant No.~NSF~PHY~1748958 at KITP.
}


\clearpage 

\section{Supporting Information}

\renewcommand\thefigure{S.\arabic{figure}}    
\setcounter{figure}{0} 

\renewcommand\theequation{S.\arabic{equation}}
\setcounter{equation}{0}

\noindent \textbf{Details of numerical approaches}

The stepping stone simulations (see, e.g., Fig.~\ref{fig: rules_of_evolution}a) use a system width of $L=2000$ sites, and are evolved until the front has advanced a height $h=1000$ sites. Results are taken from ensembles of 5000 realizations. Periodic boundary conditions are used in the direction transverse to the mean front propagation. However, in the gap and wedge geometry simulations, hard-wall boundary conditions are used, so that there is only one genetic sector boundary (instead of two), where the red sector meets the green sector.

\begin{figure}
\includegraphics[width=0.5\linewidth]{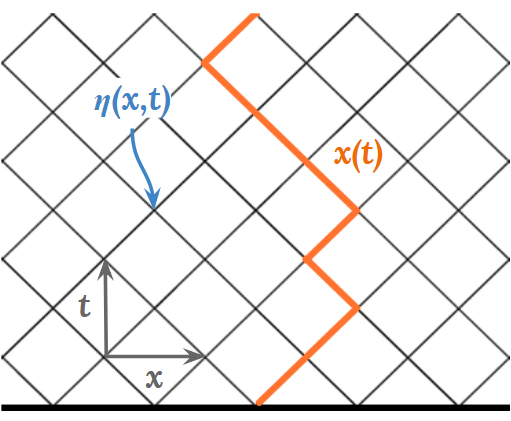} 
\caption{Schematic of DPRM on a square lattice with on-site random ``energies'' $\eta(x,t)$. As illustrated in Fig.~\ref{fig: rules_of_evolution}c of the main text, the $\eta(x,t)$ variables represent fluctuations in the cell size from generation to generation, and at different points along the $x$-axis. The path $x(t)$ propagates on average in the $t$-direction, but is allowed to wander in the $x$-direction in order to minimize the sum of random energies along the path.}
\label{fig: dprm}
\end{figure}

We simulate the DPRM (directed polymers in random media) problem on a square lattice rotated at 45$^\circ$ to the $x$, $t$ axes (see Fig.~\ref{fig: dprm}), and optimize over paths from the origin to any site $(x,t)$ using the transfer matrix method~\cite{kz87}. The simulated system has width along the $x$-direction $L=2^{16}$, is evolved over $t_{\mathrm{max}}=10^4$ time steps, and is averaged over $2^{10}$ realizations. We use periodic boundary conditions in the $x$-direction transverse to the front propagation. 

In order to avoid finite size effects, we keep the system width $L$ at least twice as large as the maximum time $t_{\mathrm{max}}$, so that no lineage or sector boundary can wind completely across the system. 

\noindent \textbf{Analytical derivation of the coalescence rate for DPRM}

Here we derive the form of the lineage coalescence rate in rough front range expansions/DPRM, Eq.~\ref{eq: J_saddlepoint}, using the DPRM endpoint distribution obtained in Ref.~\cite{fqr13}.

Consider two directed paths $x_1(\tau)$ and $x_2(\tau)$ starting from $x_1(0) = 0$ and $x_2(0) = \Delta x_0 > 0$ at $\tau=0$.
At a later time $\tau$, for $\tau/\Delta x_0^{3/2} \ll 1$, the spatial fluctuations for each path are small compared to 
their initial separation $\Delta x_0$, and we can consider the two paths to be independent.
More specifically, setting $\tilde x = x/\tau^{2/3}$,
we can take the rescaled $\tilde x_1$ and $\tilde x_2$ to be i.i.d.\ random variables drawn from the asymptotic DPRM endpoint distribution $f_{\text{end}}$ obtained in~\cite{fqr13}.
The probability distribution $f_{21}$ for the random variable $\tilde x = \tilde x_2 - \tilde x_1$ is then obtained 
from the convolution of the individual endpoint distributions, as
\begin{equation}
f_{21}(\tilde x)
= \int_{-\infty}^\infty f_{\text{end}} (\tilde y) f_{\text{end}} (\tilde y - (\Delta\tilde x_0 - \tilde x)) \; d\tilde y. \label{f21}
\end{equation}


For $\Delta \tilde x_0 \gg 1$, we are interested in the tails of the $f_{\text{end}}$ distribution, which are known to decay as $f_{\text{end}} (z) \sim \exp(-cz^3)$ with $c$ a system-specific constant~\cite{fqr13}.
This allows us to estimate the integral in Eq.~\ref{f21} using the saddle point method.
The maximum of the exponent $g(\tilde y) = c |\tilde y|^3 + c | \tilde y - (\Delta \tilde x_0 - \tilde x) |^3$ occurs at $\tilde y_* = (\Delta\tilde x_0 - \tilde x)/2$, yielding
\[ f_{21}(\tilde x) \sim \frac{\exp(-g(\tilde y_*))}{\sqrt{g''(\tilde y_*)}} \sim \frac{1}{\sqrt{\tilde x_0 - \tilde x}} \exp \left( - \frac{c}{4}(\Delta \tilde x_0 - \tilde x)^3 \right). \]

The coalescence events are represented by $\tilde x < 0$, resulting in the cumulative coalescence probability 
\begin{equation*}
C(\Delta \tilde x_0) = \int_{-\infty}^0 f_{21}(\tilde x) d\tilde x
\sim \Gamma \left( \frac{1}{6}, \frac{c\Delta \tilde x_0^3}{4} \right).
\end{equation*}
where $\Gamma(x,y)$ is the incomplete gamma function.
After properly normalizing and differentiating with respect to $\tau$, we obtain the rate of coalescence displayed in Eq.~\ref{eq: J_saddlepoint},
\begin{align}
J(\tau|\Delta x_0) \sim \frac{1}{\tau} \left( \frac{\Delta x_0^{3/2}}{\tau} \right)^{1/3} \exp \left( - \frac{c \Delta x_0^3}{4\tau^2} \right). \nonumber  
\end{align}  \\

\noindent {\bf Scaling of expected time to coalesce $T_2$}

\begin{figure}
\includegraphics[width=\linewidth]{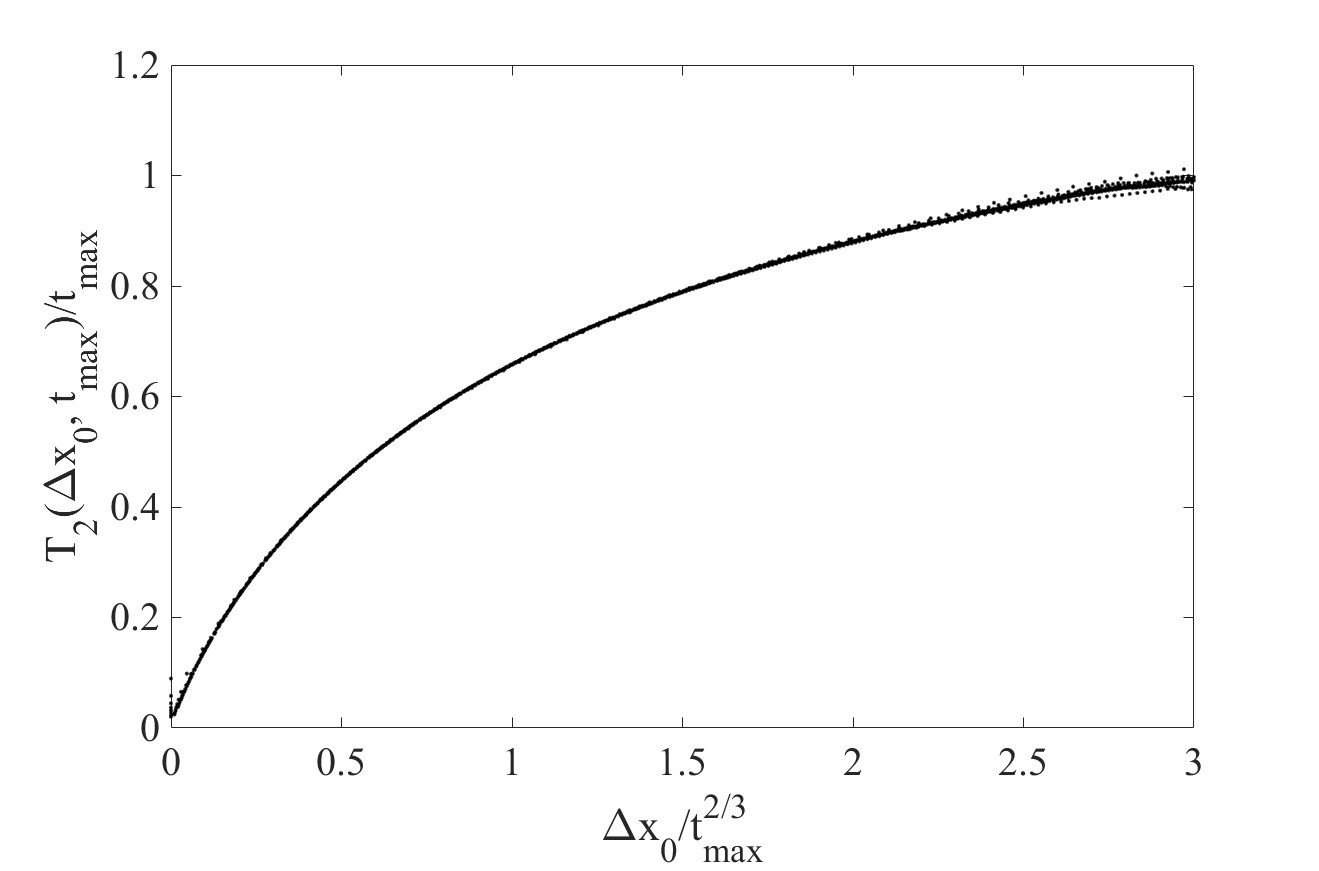}
\caption{Expected time to coalesce $T_2$ for KPZ lineages with initial separation $\Delta x_0$, collapsed with respect to the transverse scaling $\Delta x_0 \sim t_{\max}^{2/3}$. The lineages are taken from rough front stepping stone simulations of size $t_{\max} = 100$ to $1000$.}
\label{fig: T2_collapse}
\end{figure}

Analogous to the diffusive case given by Eq.~\ref{eq: T2_diff}, the expected time to coalesce $T_2$ for KPZ walkers can be written in the form
\begin{equation}\label{eq: T2_KPZ}
    \frac{T_{2,\text{KPZ}} (\Delta x_0, t_{\max})}{t_{\max}} \propto f\left(\frac{\Delta x_0^{3/2}}{t_{\max}}\right),
\end{equation}
where $f$ is some scaling function which depends only on the combination $\Delta x_0^{3/2}/t$, thus reflecting the KPZ wandering. To make this scaling relation evident, we plot a high quality collapse of the data from Fig.~\ref{fig: T2} in Fig.~\ref{fig: T2_collapse}.

\addvspace{1em}

\noindent {\bf Boundary fluctuations in the wedge geometry}

Here we present a heuristic justification of the smooth increase in the wandering exponent $\zeta$ from 1/3 to 2/3 in the wedge geometry, as the wedge angle $\theta$ is increased from $0$ to $\pi$.

Consider a wedge of opening angle $\theta$, with two distinct genotypes inoculated at its edges.
In the case of {\it flat front growth} with velocity $u$, the advancing fronts meet at a tip which zips away
from the initial apex as $y(t)=ut/\sin(\theta/2)$.
With {\it rough front growth} the sector boundary is no longer straight but meanders as the intersection of
the advancing fronts is no longer deterministic. At a time $t$, fluctuations of the front position are governed by
KPZ scaling, growing as $t^{1/3}$. While on average the time for the tip to move a distance $y$ behaves
as $y\sin(\theta/2)/u$, the fluctuations in this time scale as $[y\sin(\theta/2)/u]^{1/3}$. 

\begin{figure}
\includegraphics[width=\linewidth]{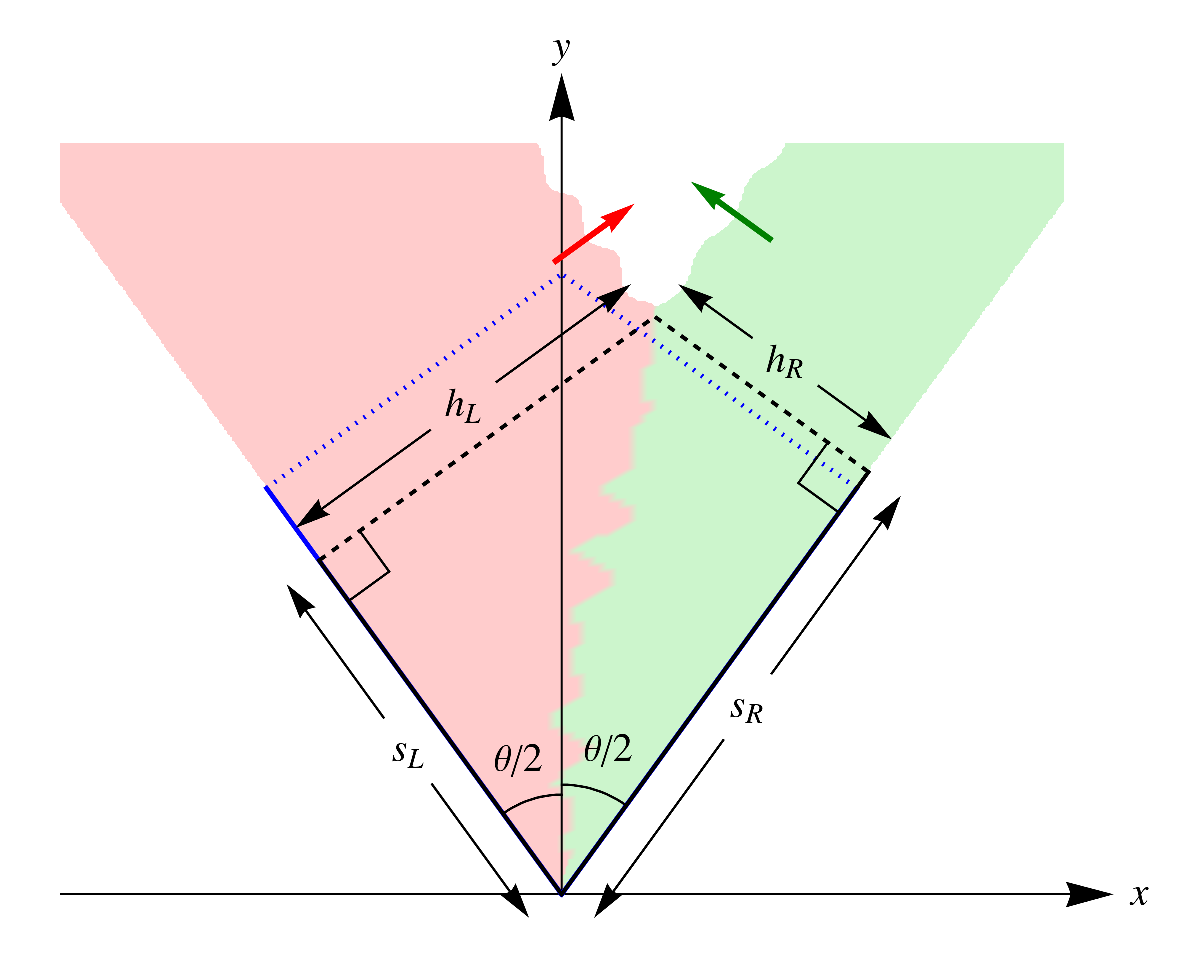}
\caption{Illustration of fluctuations in the wedge geometry with opening angle $\theta$. The red (left) and green (right) sectors meet at a sector boundary whose advancing tip, the intersection of the two dashed black lines, is pushed away from $x=0$ by fluctuations in the front propagation heights $h_L$, $h_R$, which  grow as $t^{1/3}$. The fainter blue dotted lines illustrate the zero-noise case (flat front). Coordinates $s_L$ and $s_R$ are defined to be orthogonal to $h_L$ and $h_R$, respectively.}
\label{fig: wedge_coords}
\end{figure}

The geometry is sketched in Fig.~\ref{fig: wedge_coords}. Height fluctuations $\delta h_L$, $\delta h_R$ push the advancing tip of the sector boundary -- the intersection of the black dashed lines lines -- away from $x=0$, which is the zero-noise result illustrated by the intersection of the fainter blue dotted lines. From Fig.~\ref{fig: wedge_coords}, we can solve for the intersection point $(x(t),y(t))$ representing the advancing tip:
\begin{align*}
	x(t) &= -s_L \sin(\theta/2) + h_L \cos(\theta/2)  = s_R \sin (\theta/2) - h_R \cos(\theta/2) \\
	y(t) &= s_L \cos(\theta/2) + h_L \sin(\theta/2) = s_R \cos(\theta/2) + h_R \sin(\theta/2)
\end{align*}
The height fluctuations $\delta h_L$, $\delta h_R$ can thus be expressed in terms of the resulting displacements $\delta x$, $\delta y$ of the tip, as
\begin{align*}
	\delta h_L &= \delta x \cos(\theta/2) + \delta y \sin(\theta/2), \\
	\delta h_R &= - \delta x \cos(\theta/2) + \delta y \sin(\theta/2),
\end{align*}
from which we obtain
\begin{align*}
	\delta x &= \frac{\delta h_L - \delta h_R}{2 \cos(\theta/2)}.
\end{align*}
Both $\delta h_L$ and $\delta h_R$ scale as $u t^{1/3}$, which at a given $y$ value is $u [y \sin(\theta/2)/u]^{1/3}$. Therefore, the fluctuations in $x(t)$ for a given $y$-value of the tip vary as 
\[ \delta x \propto \frac{u}{\cos(\theta/2)}\left(\frac{y\sin(\theta/2)}{u}\right)^{1/3}.
\]
While the meandering exponent remains as $\zeta=1/3$, the overall amplitude increases with $\theta$,
diverging as the wedge opens up to a single flat edge for $\theta\to\pi$. In that limit, the transverse fluctuations $\delta x$ scale as $t^{2/3}$. \\
\\

\end{document}